\documentstyle[epsf]{mn}
\newif\ifAMStwofonts

\def\gtorder{\mathrel{\raise.3ex\hbox{$>$}\mkern-14mu
             \lower0.6ex\hbox{$\sim$}}}
\def\ltorder{\mathrel{\raise.3ex\hbox{$<$}\mkern-14mu
             \lower0.6ex\hbox{$\sim$}}}

\ifoldfss
  \ifCUPmtlplainloaded \else
    \NewTextAlphabet{textbfit} {cmbxti10} {}
    \NewTextAlphabet{textbfss} {cmssbx10} {}
    \NewMathAlphabet{mathbfit} {cmbxti10} {} 
    \NewMathAlphabet{mathbfss} {cmssbx10} {} 
  \fi
  \ifAMStwofonts
    \ifCUPmtlplainloaded \else
      \NewSymbolFont{upmath} {eurm10}
      \NewSymbolFont{AMSa} {msam10}
      \NewMathSymbol{\upi}     {0}{upmath}{19}
      \NewMathSymbol{\umu}     {0}{upmath}{16}
      \NewMathSymbol{\upartial}{0}{upmath}{40}
      \NewMathSymbol{\leqslant}{3}{AMSa}{36}
      \NewMathSymbol{\geqslant}{3}{AMSa}{3E}

      \let\leq=\leqslant 
      \let\geq=\geqslant 
    \fi
  \fi
\fi 

\ifnfssone
  \newmathalphabet{\mathit}
  \addtoversion{normal}{\mathit}{cmr}{m}{it}
  \addtoversion{bold}{\mathit}{cmr}{bx}{it}
  \newmathalphabet{\mathbfit} 
  \addtoversion{normal}{\mathbfit}{cmr}{bx}{it}
  \addtoversion{bold}{\mathbfit}{cmr}{bx}{it}
  \newmathalphabet{\mathbfss} 
  \addtoversion{normal}{\mathbfss}{cmss}{bx}{n}
  \addtoversion{bold}{\mathbfss}{cmss}{bx}{n}
  \ifAMStwofonts
    \ifCUPmtlplainloaded \else
      \UseAMStwoboldmath
      \makeatletter
      \new@mathgroup\upmath@group
      \define@mathgroup\mv@normal\upmath@group{eur}{m}{n}
      \define@mathgroup\mv@bold\upmath@group{eur}{b}{n}
      \edef\UPM{\hexnumber\upmath@group}
      \new@mathgroup\amsa@group
      \define@mathgroup\mv@normal\amsa@group{msa}{m}{n}
      \define@mathgroup\mv@bold\amsa@group{msa}{m}{n}
      \edef\AMSa{\hexnumber\amsa@group}
      \makeatother
      \mathchardef\upi="0\UPM19
      \mathchardef\umu="0\UPM16
      \mathchardef\upartial="0\UPM40
      \mathchardef\leqslant="3\AMSa36
      \mathchardef\geqslant="3\AMSa3E

      \let\leq=\leqslant 
      \let\geq=\geqslant 
    \fi
  \fi
\fi 

\ifnfsstwo
  \DeclareMathAlphabet{\mathbfit}{OT1}{cmr}{bx}{it}
  \SetMathAlphabet\mathbfit{bold}{OT1}{cmr}{bx}{it}
  \DeclareMathAlphabet{\mathbfss}{OT1}{cmss}{bx}{n}
  \SetMathAlphabet\mathbfss{bold}{OT1}{cmss}{bx}{n}
  \ifAMStwofonts
    \ifCUPmtlplainloaded \else
      \DeclareSymbolFont{UPM}{U}{eur}{m}{n}
      \SetSymbolFont{UPM}{bold}{U}{eur}{b}{n}
      \DeclareSymbolFont{AMSa}{U}{msa}{m}{n}
      \DeclareMathSymbol{\upi}{0}{UPM}{"19}
      \DeclareMathSymbol{\umu}{0}{UPM}{"16}
      \DeclareMathSymbol{\upartial}{0}{UPM}{"40}
      \DeclareMathSymbol{\leqslant}{3}{AMSa}{"36}
      \DeclareMathSymbol{\geqslant}{3}{AMSa}{"3E}

      \let\leq=\leqslant 
      \let\geq=\geqslant 
    \fi
  \fi
\fi 

\ifCUPmtlplainloaded \else
  \ifAMStwofonts \else 
    \def\upi{\pi}
    \def\umu{\mu}
    \def\upartial{\partial}
  \fi
\fi

\title{A Survey for Large Separation Lensed FIRST Quasars}

\author[E.O. Ofek et al.]
       {Eran O. Ofek\thanks{e-mail: eran@wise.tau.ac.il}$^{1}$, Dan Maoz$^{1}$, Francisco Prada$^{2}$, Tsafrir Kolatt$^{3}$, Hans-Walter Rix$^{4}$ \\
$^{1}$ School of Physics and Astronomy and Wise Observatory, Tel Aviv University, Tel Aviv 69978, Israel \\
$^{2}$ Centro Astronomico Hispano-Aleman, Apdo 511, E-04080 Almeria, Spain \\
$^{3}$ Racah Institute for Physics, The Hebrew University, Jerusalem 91904, Israel \\
$^{4}$ Max-Planck-Institut f\"{u}r Astronomie, K\"{o}nigstuhl 17, D-69117 Heidelberg, Germany}

\date{Accepted ?
      Received ?
      in original form ?}

\begin{document}

\maketitle

\begin{abstract}

Little is known about the statistics of gravitationally lensed quasars
at large ($7''-30''$) image separations, which probe masses
on the scale of galaxy clusters. 
We have carried out a survey for gravitationally-lensed objects,
among sources in the FIRST 20cm radio survey that have unresolved optical
counterparts in the digitizations of the Palomar Observatory Sky Survey.
From the statistics of ongoing surveys that search for quasars
among FIRST sources, we estimate that
there are about
$9100$
quasars in this source sample,
making this one of the largest lensing surveys to date.
Using broad-band 
imaging, we have isolated all
objects with double radio components separated by $5''-30''$,
that have unresolved optical
counterparts with similar $BVI$ colours.
Our criteria for similar colours conservatively 
allow for observational error and for colour
variations due to time delays between lensed images.
Spectroscopy of these candidates shows that none of the pairs
are lensed quasars.
This sets an upper limit ($95\%$ confidence) on the lensing
fraction in this survey of $3.3\times 10^{-4}$,
assuming
$9100$
quasars.
Although the source redshift distribution is poorly known,
a rough calculation of the expected lensing frequency and
the detection efficiencies and biases suggests that
simple theoretical
expectations are of the same order of magnitude
as our observational upper limit.
Our procedure is novel in that our exhaustive search
for lensed objects does not require prior identification 
of the quasars in the sample as such.
Characterization of the FIRST-selected quasar population will
enable using our result to constrain quantitatively the mass properties of clusters.
\end{abstract}


\section{Introduction}

The statistics of gravitational lensing can provide a powerful probe
of the geometry and the mass content of the universe
out to large redshifts (e.g. Refsdal 1964; Press \& Gunn 1973). 
Turner, Ostriker, \& Gott (1984) first explored
lensing probabilities due to galaxies, and the resulting
image separation distributions.
The {\it Hubble Space Telescope}
({\it HST}) Snapshot Survey for lensed quasars (Bahcall et al. 1992; Maoz et al. 1992; 1993a;
1993b) was the first such large survey of a well-defined sample of $498$ quasars.
Exploiting the angular resolution of {\it HST}, it was shown that about 1\%
of luminous quasars at $z>1$ are gravitationally lensed into multiple
images with separations in the $0.''1-7''$ range. Maoz \& Rix (1993) used 
the Snapshot Survey results to demonstrate that early-type galaxies must
have, on average, dark massive haloes similar to those of spiral galaxies,
and that the geometry of the Universe is not dominated by a cosmological
constant $\Omega_{\Lambda}$, setting an upper limit of $\Omega_{\Lambda} < 0.7$, with
$95\%$ confidence level (CL). 
Ground-based surveys of $360$ additional quasars and their analysis
 (see Kochanek 1996, and references
therein) have confirmed these results.
Fukugita \& Peebles (1993) and
Malhotra, Rhoads, \& Turner (1997) have suggested that
small-separation lensing statistics
can be reconciled with a $\Omega_{\Lambda}$-dominated Universe
(as recently implied by high-$z$ Ia supernovae; Riess et al. 1998;
Perlmutter et al. 1999) by invoking
dust in the lensing galaxies. The excess number of lensed
quasars would then be hidden by extinction
(see, however, Falco et al. 1999).
Recently, Chiba \& Yoshii (1999) 
have recalculated
the lensing statistics,
by using revised values for the galaxy luminosity function
parameters, and have argued
that a universe
with matter density $\Omega_{0}=0.3^{+0.2}_{-0.1}$ and
$\Omega_{0}+\Omega_{\Lambda}=1$ is the most likely one.

While the statistics of gravitationally lensed quasars with multiple
images in the angular range expected due to galaxy lensing have been
probed by  the Snapshot and other surveys, less is known about lensed-quasar
 statistics at larger image separations, which probe masses
on the scale of galaxy clusters.
 Kochanek, Falco, \& Schild (1995) have reported work in progress
on a search out to $12''$ separations, but most of the large separation
candidates have yet to be rejected.
Maoz et al. (1997, see below) found no large separation lenses
among a small sample of 76 optically-selected quasars. 
The Jodrell Bank VLA
  Astrometric Survey (JVAS; King et al. 1996) and the
Cosmic Lens All-Sky Survey
(CLASS; Myers 1996) for lensing
among flat-spectrum radio sources have the potential to uncover large numbers of lensed quasars.
An extension of the JVAS
survey,
for gravitational lensing
among $\sim 2500$ flat spectrum sources in the
$6''$-$60''$ separation range, is reported preliminarily
in Marlow et al. (1998).
Phillips et al. (2000b) report a search in the
$6''$ to $15''$ separation range of the combined JVAS/CLASS sample,
with all but one
of the $\sim 15,000$ sources
currently rejected as being gravitationally lensed.
Shanks et al. (2000) discuss preliminary results for the first $6000$
quasars from the 2dF survey, in which one
large-separation candidate gravitational lens has been found.
At present, there are no confirmed cases of quasar
splittings with separations above $7''$.

Recently, Phillips, Browne, \& Wilkinson (2000a) have presented
null results from a survey for large separation lensing which has several
analogs to the survey we will describe in the present paper.
They have searched for lensed objects among
$1023$ extended radio sources in the FIRST radio survey that are
brighter than $35$mJy, and that have point-like APM optical counterparts
(See \S 2, below, for details and references for these radio and optical catalogs).
Each such radio source was searched for radio companions
brighter than $7$mJy with separations
in the range $15''$ to $60''$. 
Followup observations with the VLA and MERLIN
shows that none of the $38$ candidates
is a gravitational lens system, based on a morphological distinction
between lensed objects and physical multiple radio structures.
The choice of studying only extended FIRST sources was meant to deal with 
the problem of variability and the long time delay (tens to hundreds of years)
expected between images in 
large-separation lenses; significant changes cannot occur on such timescales
in the shapes and fluxes of the physically-extended structures in the sample
Phillips et al. (2000a)
 have defined. 

As described in detail below, our lensing survey
is limited to the $5''$ to $30''$ separation range, but uses
all FIRST sources, both extended and unresolved, that have  optical  
counterparts, and any or all of the multiple FIRST sources may be as faint as 
1~mJy. Our sample is therefore considerably larger than
Phillips et al.'s (2000a), and we show that it 
includes about 9100 quasars. As opposed to a morphological analysis in the 
radio band, our survey uses optical color criteria and spectroscopy to
reject objects as lensed. We rely on the observed 
longterm color-variation properties of quasars to address the 
time delay problem.  Our survey thus complements and extends the Phillips
et al. (2000a) work. Their null result suggests that our study has not missed
a large population of lensed quasars due to quasar 
color variability combined with
large time delays. However, the comparison of the two studies 
is not straightforward, since their different selection criteria 
result in different 
 source redshift distributions, detection limits for lensed 
pairs, and magnification biases.

The theoretically-expected
statistics of large-separation lensing have been studied by
Narayan \& White (1988), Cen et al. (1994), Wambsganss et al. (1995),
Kochanek (1995), Flores \& Primack (1996), Maoz et al. (1997),
Wyithe, Turner \& Spergel (2000),
 Li \& Ostriker (2000), and Keeton \& Madau (2000).
Maoz et al. (1997) presented the results
of a preliminary survey for
large-separation lensed quasars among known optically-selected quasars. 
They used multi-colour
photometry and spectroscopy to show that none of the point sources in the entire
$70''\times 70''$ 
field of view of the {\it HST} Planetary Camera exposures of $76$ quasars 
in the original Snapshot Survey could be lensed images of the quasars.
The 76 quasars were a selection that is unbiased against lensing 
out of the 498 Snapshot quasars, and their study demonstrated
that large image-separation lensing is not common.
Maoz et al. (1997) then carried out a calculation of the
expected lensing statistics for that particular sample and its observational
parameters. In addition to including effects such
as magnification bias and observational detection limits, their calculation
used a cluster mass profile that is motivated by N-body
simulations
(Navarro, Frenk, \& White 1995a, 1995b, 1996, 1997; NFW)
and observations (Carlberg et al. 1997; Bartelmann 1996).
They found that,
if $\gamma$, the power-law index relating
the scale radius $r_s$ to the total cluster mass in an NFW profile
is large enough (e.g. $\gamma=1$),
then low- and intermediate-mass clusters have
a large central density, and can lens more efficiently than the
singular and cored isothermal mass distributions that have been 
traditionally considered. Further study of the lensing effect of
the NFW profile, and its generalizations, have been carried out 
by Wyithe, Turner \& Spergel (2000),
 Li \& Ostriker (2000), and Keeton \& Madau (2000), the latter
tailored particularly to the observational results of Phillips et al. (2000b). 
 
There is thus a real possibility that large separation lenses
can be found in a large enough sample, and their existence can 
then be used to learn about cluster mass-structure, number density,
and redshift evolution.
Alternatively, a null result can be used to constrain these
quantities.
The mass profile of clusters is closely
related to the nature of the dark matter of which
clusters are composed, while the number density and evolution are
sensitive to the cosmological background parameters.
These considerations can be used to obtain constraints on $\Omega_{\Lambda}$, 
similar to those derived from the statistics of small separation images.
The dust extinction argument (e.g. Malhotra, Rhoads, \& Turner 1997)
is probably not applicable to
lensing by clusters, since rich clusters do not significantly redden
quasars that are behind them (Maoz 1995).

We have carried out a survey for large-separation lensed images
using a very large extragalactic source database, the FIRST
radio survey.
In \S \ref{FIRST} 
we review the basic properties of the FIRST
survey
and the APM and USNO digitizations of the first epoch
Palomar Observatory Sky Survey (POSS-I),
and in \S \ref{Sample} we describe how they were used to select
candidate lensed sources. Our observations are
described in \S \ref{observations},
and we summarize the survey
results in \S \ref{Summary},
where we also provide a rough comparison of our results to
theoretical expectations.
To interpret thoroughly the survey results we need to know
the redshift distribution
of the sources and their luminosity function, in addition to
the observational detection efficiencies.
Since not all the statistical properties of the quasars
in our sample are yet known, we postpone
the full theoretical interpretation of our results
to a future paper.

\section{Candidate Selection}
\label{sel_crit}

A survey for strong lensing requires a large,
systematically selected,
high-redshift sample of point sources.
The quasars in the FIRST radio survey constitute such a sample.
As we show below, there are $\sim 10^{4}$ quasars at redshift $z\sim 1$
in the FIRST catalog.
However, most of these are as of yet not identified.
Our strategy is, therefore, to select from the FIRST catalog
with the aid of the APM and USNO-A optical catalogs, all objects
that could potentially be quasars that have been lensed into
multiple images with large separations.

\subsection{The FIRST Radio Catalog and its Quasar Content}
\label{FIRST}

FIRST, {\it Faint Images of the Radio Sky at Twenty-cm} (Becker, White \& Helfand 1994, White et al. 1997)
is a project designed to produce the radio equivalent of the
POSS over $10,000\Box^{\circ}$ of the north and
south Galactic caps.
The survey utilizes the National Radio Astronomical Observatories (NRAO)
Very Large Array (VLA) in
the B configuration with
bandwidth-synthesis
mode and fourteen 3-MHz-wide channels centered at 1400 MHz (20 cm).
This allows 
for a relatively
small beam size ($5.''4$ in the northern catalog and $5.''4\times 6.''4$
in the southern catalog) and hence good angular resolution and
astrometric accuracy
($90\%$ confidence positional error circle better than $1''$).
The beam size
enables the detection of source structure down to scales of $\sim
2.''5$.
Comparisons between the FIRST catalog and standard radio calibration
sources, indicates that the systematic astrometry errors are smaller
than $0.''2$ (Gregg et al. 1996).
The typical root-mean-square (RMS) noise of 0.15 mJy allows $5 \sigma$
detection of 1 mJy sources.

For our work, we used
the 1999, July 21, version of the FIRST catalog, including
$\sim 5400 \Box^{\circ}$,
above Galactic latitude $+25^{\circ}$, between declinations $-5^{\circ}$
and $+58^{\circ}$. The catalog also covers two
narrow strips in the southern Galactic cap
centered near declinations $0^{\circ}$ and $-9^{\circ}$, comprising an
additional $\sim 610 \Box^{\circ}$.
There are
$549,707$ sources
in the entire catalog, of which
$54,537$ are in the southern catalog.
Over 99.9\% of the sources in the FIRST catalog are
extragalactic (Helfand et al. 1999).
Helfand et al. (1998b) estimate that the mean redshift of FIRST
sources is  $z \sim 1$.

The FIRST Bright QSO Survey (FBQS; Gregg et al. 1996; White et al. 2000),
is a spectroscopic
survey of all FIRST (1997, April 24 version, $2682 \Box^{\circ}$) objects with
optical counterparts within $1.''2$ of the FIRST sources
that are classified as stellar
on either of the two emulsions, $O$ (blue) or $E$ (red)
in the Automated Plate Machine (APM; McMahon, \& Irwin 1992)
scan of the POSS-I/UKST plates,
that after correcting for Galactic extinction,
are brighter than $E_{APM}=17.8$~mag,
and have colour $O-E<2$~mag.
The $O$ and $E$ passbands have effective wavelengths (widths)
of roughly
$4200$\AA ($1200$\AA) and $6400$\AA, ($400$\AA), respectively.
White et al. (2000) found that $51.4\% \pm 2.0\%$ (Poisson errors)
of FIRST sources passing their selection criteria are quasars.
The preliminary FBQS sample described by Gregg et al. (1996),
used a positional coincidence
criterion of $2.''0$ instead of $1.''2$, and no colour criterion.
They showed that
the more strict positional criterion would
eliminate only about $4.4\% \pm 2.5\%$ (Poisson error) of the
FBQS quasars.
However, the requirement of close positional coincidence excludes
extended, lobe-dominated, radio sources with no core
component.
Although Gregg et al. (1996) did not find any red quasars with
$O-E\geq2$~mag,
the red colour cut used in the full FBQS sample may make the survey
incomplete for high redshift
or obscured quasars.
The mean and median redshifts of FBQS quasars are 1.05 and 0.95, respectively.

Becker et al. (1998) described an extension to the FBQS,
to a limiting magnitude of $E_{APM}=19.0$~mag, named the
FIRST Faint Quasar Survey (FFQS).
The other FFQS selection criteria are similar to those of the FBQS.
The fraction of quasars in the FFQS
is about $90\%$, almost all with $z>0.5$
(M. Brotherton, 1999 - private communication).
The FFQS sample incompleteness may be greater due
to the colour cut, since the fainter quasars may be
at higher redshift and hence  redder.
However, the number of quasars missed in the FFQS
because of the colour criterion is unknown. 
Helfand et al. (1998a) describe a search for optical counterparts
to radio sources from the FIRST survey, using the deep $16 \Box^{\circ}$ 
$I$-band survey of Postman et al. (1997) to a limiting magnitude
$I\sim 24$. They detect $700$ out of $1131$ FIRST sources in this field.
Spectroscopic identifications have been obtained for a significant fraction
of the stellar counterparts. Most, as expected, are quasars.

An independent digitization of the POSS-I and UKST plates
is the USNO catalog
(Monet et al. 1997), which covers the entire sky.
It includes all sources that
have positional coincidence
to within $2''$
on both the
$E$ and $O$ plates (in the northern hemisphere), or
the $SRC-J$ and $ESO-R$
plates (in the southern hemisphere) of the POSS-I/UKST.

Our survey for
large separation lensed quasars
utilizes both the APM and the USNO-A1.0 catalogs
to produce two candidate lensed quasar samples,
as described in details below.
In order to relate these two digitizations of the
POSS-I, we compared the photometry of the
USNO-A catalog, the APM photometry,
and calibrated Johnson-Cousins $B$ and $R$ photometry of stars
in random fields measured at Wise Observatory.
The mean scatter in the USNO-A1.0 photometry
is $0.38$~mag in the $E$ band, and $0.36$~mag
in the $O$ band.
In some magnitude ranges the USNO magnitudes also have
systematic offsets of $\pm 0.3$~mag.
For the APM catalog we find a smaller scatter of up to
$0.2$~mag and systematic errors of up to $0.7$~mag.
The APM $E=17.8$~mag limit
in the FBQS corresponds to $E \approx 18.0$ in the
USNO-A1.0 magnitude system and it is equivalent to $R \approx 17.8$~mag
in the Cousins R band.
The $E=19.0$~mag limit of the FFQS corresponds to $E \approx 19.0$~mag
in the USNO-A1.0 magnitude system
and to $R\approx 19.0$~mag in the Cousins system.
Caretta et al. (2000) compared the APM catalog with several others
deep surveys, among them the ESO Imaging Survey
(Nonino et al. 1999). They found
that the APM is $\sim 100\%$ complete to
$O_{APM}=19.5$~mag, and $\sim 70\%$ complete to
$O_{APM}=21.5$~mag.

\subsection{Source Samples and Selection of Lensed Candidates}
\label{Sample}

We create a large extragalactic source sample, which we survey
for large-separation lensed quasars, by correlating the
FIRST catalog with the APM and USNO-A1.0 catalogs,
as described below. The APM catalog
covers about $96\%$ of the area covered by the FIRST catalog,
while the USNO-A1.0 catalog covers the entire sky.
To produce our candidate lensed quasars for the
FIRST-APM sample, we searched for APM
optical counterparts within $2.''5$ of all FIRST
sources\footnote{The catalog of APM optical counterparts is available from: $http://wise-obs.tau.ac.il/\sim eran/SLSL/$}.
We found $86,800$ optical counterparts within $2.''5$,
and $64,154$ optical counterparts within $1.''2$, of the radio position.
Although the FIRST
90\%-confidence error radius is less than $1''$, we chose the large
threshold to ensure that
we do not miss any optical counterparts.
Furthermore, some of the quasars could have extended radio structure
that is not completely coincident with the optical source.
While this large threshold increases the number of candidate lenses
that need to be tested by subsequent observations,
it does not adversely affect the statistics of the survey.

Following the FBQS selection criteria,
we isolated all the pairs of radio-optical sources having separations of $5''$
to $30''$, in which both pair members are point-like in
at least one of the POSS/UKST $O$ or $E$ plates and in which both members
have $O-E<2$~mag.
As in the FBQS, before implementing the colour and magnitude
criteria, we applied an extinction correction to each object
using the $E(B-V)$ map of Schlegel, Finkbeiner, \& Davis (1998),
$A(E)=2.7E(B-V)$ and $A(O)=4.4E(B-V)$.
These corrections are usually quite small.
The median values for our source list
is $A(E)=0.068$~mag and $A(O)=0.111$~mag.

In order to estimate the number of quasars in the FIRST-APM sample,
we have counted the number of FIRST optical counterparts in the APM
catalog satisfying the 
FBQS/FFQS criteria (point-like in at least one of the plates,
$O-E<2$~mag, and up to $1.''2$ positional coincidence),
as a function of magnitude.
We found $2155$
optical counterparts brighter than $E_{APM}=17.8$~mag
and $8300$
optical counterparts fainter than $E_{APM}=17.8$~mag,
or a total of $10,455$ objects with
up to $1.''2$ positional coincidence
(or $12,576$ objects with up to $2.''5$ positional coincidence).
We fit a second order polynomial to the observed 
fraction of quasars among FBQS candidates, as a function of
magnitude,
given in Figure~4 of White et al. (2000). 
Between  $E_{APM}=18$ and $19$~mag, the fit 
is constrained  to a fraction of 90\%, in accordance
with the preliminary FFQS results
(M. Brotherton, 1999 - private communication).
Among optical counterparts
fainter than $E_{APM}=19$~mag, we assumed the same quasar fraction
fraction as in
the FFQS. Our 
adopted fraction of quasars ($F_{qso}$) as function of
$E$ magnitude is:
\begin{equation}
F_{qso} = \left\{ \begin{array}{ll}
          -0.9389 + 0.046E + 0.0027E^{2}, & \mbox{$E<18$} \\
          0.9,                            & \mbox{$E\geq 18$}
          \end{array} \right.
\label{Frac_qso}
\end{equation}
The number of quasars in our sample is given by
integrating, over magnitude, the fraction of quasars multiplied by
the number of candidates, $N_{cand}(E)$, as a function of magnitude:
$\int{F_{qso}(E) N_{cand}(E)}dE \cong 8900$.
As mentioned above, Gregg et al. (1996)
found that about $4\%$ of the optical counterparts
with a radio-optical positional coincidence in the range
$1.''2$ to $2.''0$ are quasars, adding another
$\sim 200$ quasars to our sample. 
Thus the number of quasars
in our sample, of the type being found by the FBQS and FFQS,
is about
$9100^{+500}_{-4000}$,
where the
bounds are obtained by assuming the
fraction of quasars among
objects fainter then $E_{APM}=19$~mag is 0 or 1.
However, from the preliminary
results of Helfand et al. (1998a),
it is unlikely that the number of quasars in our survey
is near the lower bound (e.g. $5100$).
A more accurate estimate of the number of quasars
in the survey and their properties must await
the full results of the FFQS and the optical
identifications of radio sources in the Postman et al. (1997) field.
We are also exploring the use of 
the Sloan Digital Sky Survey (SDSS) to obtain
a better estimate of the faint quasars fraction.

The procedure described above could potentially exclude
some FIRST quasars from our source list.
The colour criterion used in the FBQS and FFQS
could exclude a population of 
red quasars (e.g., quasars with $z\gtorder 2.5$, or highly
obscured quasars).
Based on the preliminary FBQS (Gregg et al. 1996),
the fraction of missed red quasars is probably small in the
FBQS ($<4.3\%$ with $95\%$ CL; from Poisson statistics),
but possibly larger in the FFQS.
Some FIRST quasars could also be missed if the APM catalog
misclassified some point-like sources as galaxies.
However, this is less likely since Caretta et al. (2000)
found that the fraction of point-like objects that were
misclassified as galaxies by the APM
is smaller than $3\%$, up to $O\sim 20.5$.
Finally, either of the two
POSS digitizations, the APM and the USNO catalogs, could be incomplete.

To investigate for the possible signatures of these effects,
we created a second source sample by cross-correlating an earlier
(i.e., smaller)
version of the FIRST
catalog (1998, February 4th; $4760 \Box^{\circ}$) with the USNO-A1.0 catalog.
In creating this sample, we ignored the POSS colour information,
i.e., we included both blue and red objects.
We also did not rely on the morphological
classification of the APM (contrary to the APM, the USNO catalog does
not provide morphological information), 
but rather accepted objects as point-like
only based on our own subsequent CCD images. 
Thus, this sample may be considered a conservative sub-sample,
intended to test for the potential problems outlined above.
For each pair of FIRST radio sources we used a modified version of the
{\sc REFNET}\footnote{The original
  Astronomical REFerence NETwork program was written by Ted Bowell \&
  Bruce Koehn, Lowell Observatory. ftp://ftp.lowell.edu/pub/koehn/starnet/dist.html}
program to search the USNO-A1.0 catalog
for optical counterparts to both members of the radio pair within $2.''5$.

Comparing the source lists produced with the APM and the USNO
catalogs when the same criteria are applied to both (i.e., 
no morphological or colour criteria), we find that the APM list
includes $\sim90\%$ of USNO sources, and the USNO-A1.0 catalog
includes a similar fraction of APM sources.
A check of the missing sources in both catalogs,
shows that they are extended galaxies,
or in few cases, bright stars ($E\ltorder 12$~mag)
for which the
source centering by USNO-A1.0 and APM is different.
We conclude that this test does not reveal
evidence for incompleteness in either catalogs.

The APM sample selection process produced $9$ pairs
($7$ in the north and $2$ in the south)
having up to $30''$ separation between pair members,
while the USNO sub-sample
selection process produced
$226$ pairs ($209$ in the north, $16$ in the south).
There is overlap between these two samples, such that there is
a total of 230 actual pairs.
In about $20\%$ of these pairs, at least one of the members is listed
as a galaxy in the
NED\footnote{NASA-IPAC Extragalactic Database, $http://nedwww.ipac.caltech.edu/$}
archive and was therefore eliminated
as a lensed quasar.
In principle,
the two-band POSS photometry (either APM or USNO) and the FIRST
20 cm flux could have been used to reject as lensed those objects having
different flux ratios in different bands.
However,
as shown in \S \ref{FIRST}, the POSS
magnitudes have a large error
of $\sim0.2/0.4$~mag (for the APM and USNO-A1.0 respectively). Furthermore,
the time delay expected between images in the case of large separation
lenses 
can be of order of years to centuries,
and we lack any
knowledge
on the variability of the optical-to-radio ratio on
such large time-scales.
We therefore ignored the FIRST and POSS flux/magnitude
information when compiling our candidate lists.

In summary, we have defined from the FIRST catalog a sample
of sources with the following properties:
All radio sources brighter than $1$mJy with point-like APM
optical counterparts
($\sim 70\%$ complete to $O_{APM}=21.5$~mag)
that are bluer than $O-E=2$~mag.
In a second sample used mainly for completeness checks, we used
the USNO-A1.0 catalog and a FIRST sub-sample,
this time with no colour or morphology criteria.
From both samples we have isolated all pairs with $\leq 30''$ separation,
which constitute a sample of candidate lensed quasars.

\section{Observations}
\label{observations}

In order to proceed
we carried out broad-band photometry
and/or spectroscopy for all pairs in the two samples
that are not listed as galaxies in the NED archive.
Using the Wise Observatory 1.0m telescope
with a back-illuminated Tektronix $1024\times 1024$ CCD,
on $50$ nights
between 1998, April
and 1999, July, we acquired $I$, $V$ and $B$ images
for $178$ pairs.
Some $R$-band images of the pairs
were kindly obtained by
R. Uglesich in 1999, January
with the MDM 1.4m telescope using
the
Echelle $2048\times 2048$ CCD.

The images were reduced in the standard way, using the
{\sc IRAF}\footnote{Image Reduction and Analysis Facility - is written and
  supported at the National Optical Astronomy Observatories (NOAO)}
package.
We programmed a task\footnote{Available from
  $http://wise-obs.tau.ac.il/\sim eran/iraf/$}
to identify the sources automatically
by finding the astrometric solution of each image.
The optical and radio positions for the pair were then
marked on the image for visual inspection,
and the optical source coincident with the
radio source was measured using the APPHOT task.
Sources that were resolved in any of the bands were rejected
as lensed quasars.
About $84\%$ of the pairs were rejected based on this
criterion. 
Although resolved galaxies
undergo strong
lensing with large image separations,
in that case they are always distorted into arcs.
Our point source criterion
might however, accidentally exclude multiply-imaged quasars
in which one or more nearly-merging image pairs appear as one,
marginally-resolved object,
and this will need to be accounted for in the theoretical
interpretation of our results.

Next, we measured the flux ratios in each band, and compared
them.
Since gravitational lensing is achromatic,
pairs with different flux ratios
(see below) were rejected
as being lensed quasars.
A possible problem with the flux-ratio comparison is the time delay
between lensed images.
For cluster lenses, this delay can be of order years to centuries.
Giveon et al. (1999) monitored $42$ ($35$ radio-quiet and $7$ radio-loud)
quasars for $7$
years in the $B$ and $R$ bands, and found that the $1\sigma$ $B-R$ variability
in their sample was $0.053$~mag.
In order to minimize false rejections,
we allowed for such variability when
calculating the flux ratios of pairs.
It is possible that on time-scales longer than
$\sim 7$
years, but which are still relevant to large
separation lensing, larger colour changes occur in quasars.
If so, our flux-based rejection criterion
(and even spectroscopic criteria)
may eliminate true lensed pairs.
However, Helfand et al. (2000) recently examined the $B$ and $R$ 
variations of FIRST quasars over $\sim 50$ yr timescales, by comparing
CCD photometry to the APM plate-based magnitudes. Their results suggest 
a small typical color change, comparable to that measured by Giveon et al.
(1999), over these longer timescales. Nevertheless,
the possibility of color changes needs to be tested by future studies of
quasar variability,
and by studying actual large-separation systems
when such are found.

\begin{figure*}
\vspace{20cm}  
\includegraphics{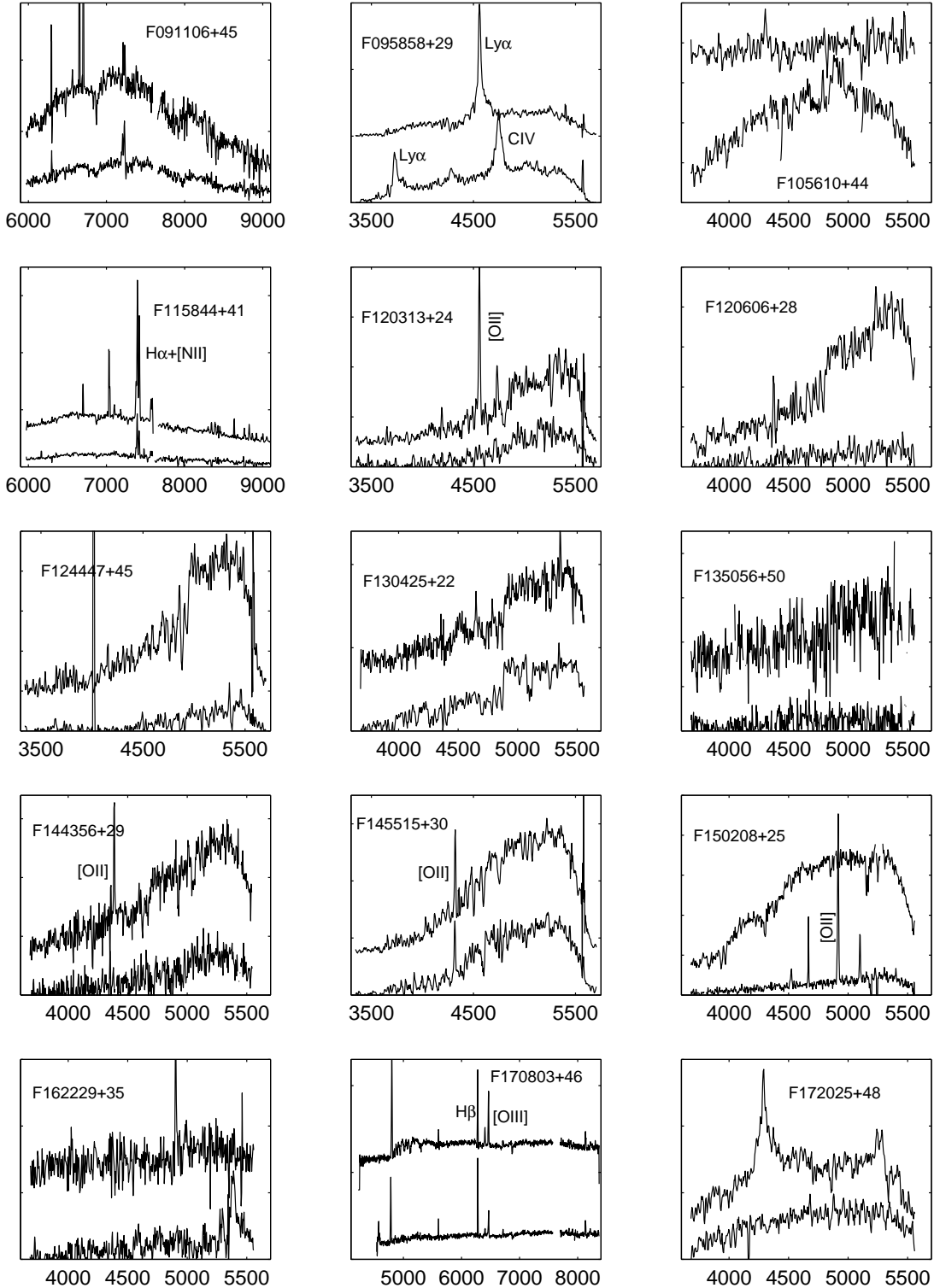}
\caption {Spectra of candidate pairs. Vertical axis is counts in
arbitrary units and horizontal axis is wavelength in \AA.
Additive shifts have been applied to some of the spectra
for display purposes.
Spectral regions with strong telluric absorption have been excluded.
All spectra are from the Calar-Alto 3.5m telescope, except for F170803+46,
 which
is from the Keck-II 10m telescope.}
\label{all_spectra}
\end{figure*}

In each band,
we calculated the magnitude difference, and its
error, between the pair of sources.
The magnitude differences in the three bands were compared
in units of the standard error, and we rejected as possible
lensing candidates the pairs for which the
flux ratio in one or more of the optical bands differs by more then
$3.5 \sigma$:
\begin{equation}
\vert \Delta m_{i} - \Delta m_{j} \vert > 3.5\sqrt{\delta^{2}\left( \Delta
    m_{i}\right) + \delta^{2}\left( \Delta m_{j} \right) + 0.053^{2}},
\label{eq:worst_dm}
\end{equation}
where $\Delta m_{i}$ is the magnitude difference between a pair
of objects in the $i-th$ filter, $\delta (\Delta m_{i})$ is the error in
the magnitude difference between a pair of objects in the $i-th$ filter,
and $0.053$~mag is the allowed $1\sigma$ colour variability.
Following this analysis,
$15$ of the pairs from the two samples
remained candidate lenses.
Table~1\footnote{The full information for all the observed
pairs can be found at $http://wise-obs.tau.ac.il/\sim eran/SLSL/$.}
lists the candidates with their coordinates,
approximate APM optical magnitude, and radio flux.
If available, for each candidate the optical magnitude differences
and the largest colour difference
(including the $B-R \approx 0.053$~mag variability) in units of $\sigma$,
are listed.
\bigskip
\begin{table*}
\vskip 0.2truein
\begin{center}
\caption{Lensed Quasar Candidates}
\begin{tabular}{c}
PLEASE PLACE TABLE 1 HERE\\
\end{tabular}
\vskip 0.2truein
\end{center}
 \end{table*}
\bigskip

We obtained spectra for all the lensed candidates in 1999, May 17, and
2000, March 6, 7 and 8.
The spectra  
were obtained using the
TWIN double-channel spectrograph on the Calar-Alto 3.5m telescope,
with the T07 \& T13 grating and the $5500$\AA ~dichroic,
covering the red channel with
$3.3$~\AA~pixel$^{-1}$ dispersion, and the blue
channel with $4.3$~\AA ~pixel$^{-1}$ dispersion, respectively.
The $1.''8\times 240''$
spectrograph slit was rotated to allow simultaneous observations of
both members of each object pair.
The spectra for additional pair was obtained on 1998, August 15,
using the Keck-II 10m telescope with the LRIS spectrograph
and $2.45$~\AA~pixel$^{-1}$ dispersion.
Figure~\ref{all_spectra} presents the spectra of the candidates.

The pairs are clearly ruled out as lensed quasars.
All are either physical pairs of galaxies
at $z\sim 0.2$, with velocity differences of order
of $200$ km s$^{-1}$, or pairs
of unrelated galaxies and quasars at different redshifts.
Interestingly, our selection process, combined with the
apparent rarity of large-separation lensing, seems
to be successful at picking up
physical pairs (possibly belonging to poor groups)
of $z\sim 0.1-0.3$ radio galaxies.

\section{Discussion}
\label{Summary}

We have surveyed a sample of about
$9100$
radio-selected quasars 
for large separation lensing.
This is one of the largest surveys of its kind.
The survey utilized the FIRST catalog in the radio, and the APM and USNO-A1.0
catalogs in the optical.
Our observations show that none of
the candidates turned up by our selection
process is a gravitationally lensed quasar.
Assuming the number of quasars in our FIRST-APM sample is
$9100$,
this sets the probability for
lensing in our sample at 
$<3.3\times 10^{-4}$ with $95\%$ CL (assuming Poisson statistics).
This is an improvement of two-orders of 
magnitude over the large-separation lensing survey of Maoz et al. (1997).
Our observations have identified several chance superposition of
quasars and radio galaxies
and several physical pairs of radio galaxies.
We found no evidence for a missed population of red or misclassified quasars 
due to the FBQS, FFQS, or APM selection criteria, a population that could
have been detected in the FIRST-USNO sample.

Our well defined statistical sample can be used to constrain the density
profiles and mass
function of galaxy clusters and to set limits on $\Omega_{\Lambda}$
(e.g., Maoz et al. 1997). 
Although large-separation lensing
probabilities have been calculated in the past
(e.g., Flores \& Primack 1996; Maoz et al. 1997, and reference therein),
a detailed interpretation of the results reported here
requires good knowledge of (1) the joint radio-optical luminosity
function, to quantify magnification bias and
(2) the redshift distribution of the quasars in our sample,
to quantify the probed pathlength.
We therefore must defer the full calculation of the theoretically expected
lensing probability until completion of the FFQS
and the spectroscopic identification of FIRST sources in the 
Postman et al. (1997) field.

However, a rough estimate of the optical depth, an hence the
expected lensing frequency,
can be obtained under various simplifying assumptions:
We approximated the
mass distributions of clusters as singular isothermal spheres (SIS),
and then, following Turner et al. (1984),
calculated the effective dimensionless density of lenses,
$F=16\pi^{3}n_{0} \left(\frac{c}{H_{0}} \right)^{3} \left( \frac{\sigma_{\parallel}}{c} \right)^{4}$,
where $c$ is the speed of light, $H_{0}$ the Hubble constant,
$\sigma_{\parallel}$ the line-of-sight velocity dispersion,
and $n_{0}$ the density of lenses.
We assumed the observed Girardi et al. (1998) mass function for clusters
and related the cluster mass, $M$, within $r=1.5 h_{100}^{-1}$ Mpc
(as defined by Girardi et al. 1998)
to the line-of-sight velocity dispersion
by $\sigma_{\parallel}^{2} = \frac{GM(<r)}{2 r}$,
giving
$F=0.0156_{-0.0067}^{+0.0118}$.
The errors in $F$ were obtained from Monte-Carlo simulations
using the errors in the Girardi et al. (1998) mass function parameters
$n^{*}=2.6_{-0.4}^{+0.5} \times 10^{-5} (h_{100}^{-1}$ Mpc$)^{-3} (10^{14} h_{100}^{-1} M_{\odot})^{-1}$, and
$M^{*}=2.6_{-0.6}^{+0.8} \times 10^{14} h_{100}^{-1} M_{\odot}$.
For simplicity,
we assumed that the errors in $n^{*}$ and $M^{*}$ are independent and normally distributed.

\begin{figure}
\centerline{\epsfxsize=85mm\epsfbox{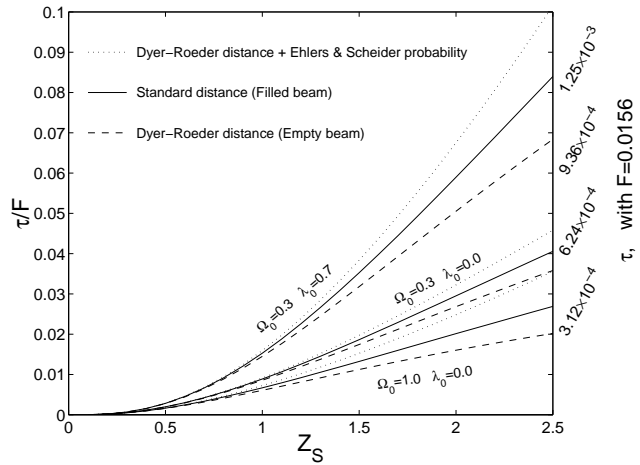}}
\caption {Weighted total optical depth for lensing by SIS clusters with
the Girardi et al. (1998) mass function as a function
of the source redshift.
Three cosmological models are labeled, with three
different distance/probability estimators for each model.
$F$ is the density of SIS lenses as defined by Turner et al. (1984).}
\label{total_cross_section}
\end{figure}

Figure~\ref{total_cross_section} presents the SIS results for three
cosmological models: 
$\Omega_{0}=0.3$, $\Omega_{\Lambda}=0.7$;
$\Omega_{0}=0.3$, $\Omega_{\Lambda}=0.0$; and
$\Omega_{0}=1.0$, $\Omega_{\Lambda}=0.0$.
For each model we calculated the optical depth with the
standard distance formula (solid line), the Dyer-Roeder distance formula
(Dyer \& Roeder 1972; 1973; broken line) and the Dyer-Roeder distance formula
with the Ehlers and Schneider probability (1986; dotted line).
For a median redshift of the FIRST quasars ($z\sim 1$) the
optical depth is about $(0.5-4.4) \times 10^{-4}$.
To this range, a factor of $3.1$ is contributed
by the uncertainty in the mass function,
and a factor of $2.3$ is contributed by the uncertainty in
the cosmology.

For an estimate of the lensing frequency, the optical depth
to lensing needs to be multiplied by the following factors:
(i) The detection efficiency due to the
limited range of searched separations;
(ii) The detection efficiency due to the survey's double-flux limit
and the expected flux-ratio distribution;
(iii) The magnification bias - the degree to which,
at a given magnitude, lensed
quasars are over/under -represented.

As described above, we searched for lensed quasars in
the $5''$ to $30''$ separation range.
Since we do not yet know
the redshift distribution of FIRST quasars,
we cannot calculate the exact image separation distribution
expected from SIS lenses.
However, assuming the typical ratio
between the lens-source distance and observer-source distance
is $D_{ls}/D_{os}=1/2$,
the fraction of lenses with separations of
$5''$ to $30''$ is $66\%$.

\begin{figure}
\centerline{\epsfxsize=85mm\epsfbox{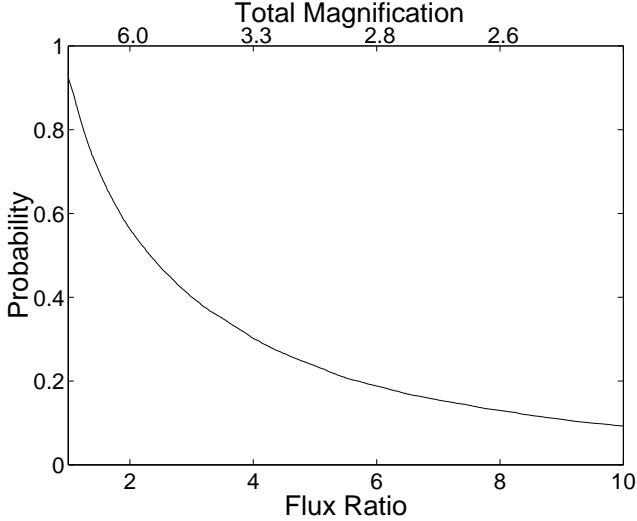}}
\caption {The probability that a lensed
image of a quasar will pass the double ($O$ and $20$cm bands)
detection limit of our survey, as a function of the flux ratio between
the two images, or equivalently, as a function of
the total magnification of a SIS lens.}
\label{FR_PDF}
\end{figure}

We have searched for lensed
systems in a doubly (radio and optical)
flux-limited sample. Since most objects are near the detection
limit, the fainter member of the pair is often below the detection limit.
In order to estimate this effect, we calculated
the probability that a lensed
image of a quasar will pass the double
detection limits of our survey in the $O$ and $20$cm bands,
as a function of the flux ratio between the two images.
This was done as follows.
From among all the optical counterparts with $O-E<2$
found within $1.''2$ from a radio source,
we chose at random a list of ''quasars'',
using Equation~\ref{Frac_qso} for the fraction of quasars
as function of magnitude.
For each object in the list,
the maximum observable flux ratio was found,
using its optical magnitude and radio flux.
The maximum observable flux ratio is defined as the flux ratio
that a hypothetical
fainter lensed image would have, and still be detected in both
optical and radio bands.
The flux ratio probability was then calculated.
Until detailed information about the FIRST
radio flux limit is available, we assumed it to be a step function at
$1$mJy.
For the APM $O$-band detection limit, we adopted
the APM completeness given by Caretta et al. (2000).
Since $98\%$ of the objects in our APM sample are brighter in
$E$ than in $O$ (i.e. $O-E>0$), and
the $E$-band completeness is similar to that of the $O$-band,
we ignored the $E$-band completeness.
This procedure assumes that
the correlation between
the radio flux and optical magnitude is negligible
(as shown below),
and that the flux ratio between the $O$ band and
the radio is the same for both images (i.e. neglecting variability).
Figure~\ref{FR_PDF} shows
the result of this calculation.

\begin{figure}
\centerline{\epsfxsize=85mm\epsfbox{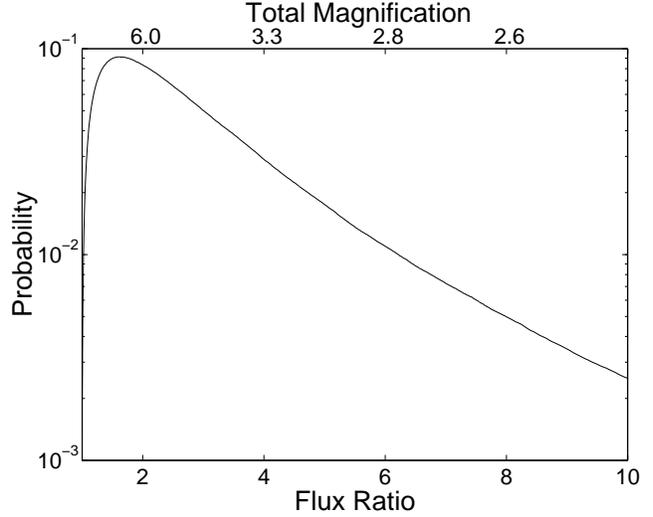}}
\caption {The probability density
for finding a doubly-imaged lensed quasar with a given flux ratio,
calculated by multiplying the flux ratio distribution of
a SIS lens by the double flux limit selection function of the survey.
The total magnification of a SIS lens is plotted in the upper axis
for reference.}
\label{FR_Prob}
\end{figure}

Next, we considered the flux ratio distribution
predicted by the lens model.
The flux ratio between the two images in a SIS lens
is $R=\frac{1+\beta}{1-\beta}$ and
the probability of an impact parameter $\beta$ is
$dP=2\beta d\beta$. The flux ratio probability is
$dP=4\frac{R-1}{\left(R+1\right)^{3}}dR$.
Figure~\ref{FR_Prob} shows the probability density as function
of the flux ratio
resulting from the product of the
intrinsic flux ratio distribution due to the lens model
and the flux ratio selection function (described above).

\begin{figure}
\centerline{\epsfxsize=85mm\epsfbox{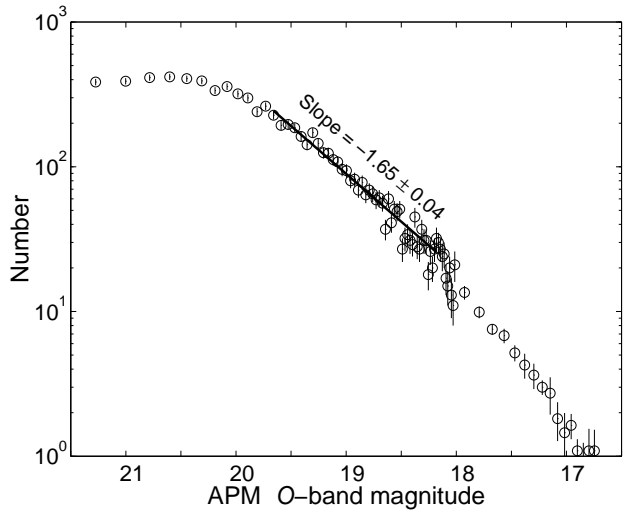}}
\caption {Differential number counts for our
sample quasars in the $O$ band.
To account for the quasar fraction, as measured in the
FBQS and FFQS,
in each magnitude bin a fraction $F_{qso}(E)$
(as given by Equation~\ref{Frac_qso}) of objects
was randomly chosen.
A power-law is fit to the data in the range $O_{APM}<19.75$~mag
($100\%$ completeness; Caretta et al. 2000)
and $O_{APM}>18.25$~mag, where a change of slope is apparent.
The slope
$-1.65\pm0.04$, corresponds to
$0.66$ in the $\log{N}$-mag plane.
The points for objects brighter than $18$~mag are the means
of 10 adjoining bins.}
\label{CumNumCountO}
\end{figure}

\begin{figure}
\centerline{\epsfxsize=85mm\epsfbox{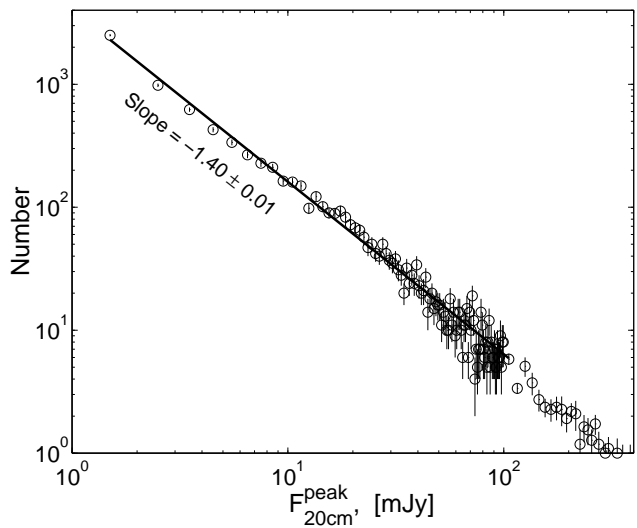}}
\caption {Same as Figure \ref{CumNumCountO}, but for the $20$cm band.}
\label{CumNumCount20cm}
\end{figure}

Following Turner et al. (1984), the magnification bias is given by,
\begin{equation}
B = \frac{\int_{0}^{\infty}{\int_{A_{M}}^{\infty}{A^{-1}P_{I}(A)P_{D}(A)N_{Q}(f/A)}dA}df}{\int_{0}^{\infty}{N_{Q}(f)}df},
\label{eq:magnification_bias}
\end{equation}
where $f$ is the flux, $A_{M}$ is the minimum
amplification of the bright image (see Mortlock \& Webster 2000)
for a multiply imaged
source (e.g. $2$ for a SIS lens), $P_{I}(A)$ is
the intrinsic probability density for the amplification $A$
of the brightest image
due to the lens model, $P_{D}(A)$ is the
detection probability for the amplification $A$
of the brightest image
due to the flux ratio detection efficiency of the survey, and
$N_{Q}(f)$ is the number of quasars
with unlensed flux $f$.
Borgeest, Linde, \& Refsdal (1991) showed that if there is
a negligible correlation between the optical and radio
fluxes of the objects,
the double optical-radio magnification bias depends on the
sum of the slopes of the radio and optical
differential number counts.
In our case the correlation
is $\sim 4\%$ with $\sim10^{4}$ degrees of freedom.
Figure~\ref{CumNumCountO} shows the binned number counts for our
sample's quasars in the $O_{APM}$ band.
Figure~\ref{CumNumCount20cm} shows the same for
the $20$cm band.
To account for the quasar fraction, as measured in the
FBQS and FFQS,
in each magnitude bin a fraction $F_{qso}(E)$
(as given by Equation~\ref{Frac_qso}) of objects
was randomly chosen.
The optical and radio slopes of the
differential number counts are
$\alpha_{opt}=-1.65\pm0.04$ with $\chi^{2}_{dof}=1.17$,
and $\alpha_{20cm}=-1.40\pm0.01$ with $\chi^{2}_{dof}=1.59$, respectively.
The total slope is $-3.05\pm0.04$.
For a slope of $-3$, or smaller,
the magnification bias integral diverges.
In reality, however, there are probably breaks in the
number count relations.
The detection of such breaks is critical for the
exact assessment of the magnification bias. 
We plan to address this problem by deriving
the number-magnitude relations of similar, but smaller,
radio-optical samples of objects.
Such samples could be constructed from deeper optical
(e.g. SDSS) and $21$cm radio surveys
(e.g. the PHOENIX Deep Survey, Hopkins et al. 1998; 
the ISO ELAIS Survey, Ciliegi et al. 1998).

\begin{figure}
\centerline{\epsfxsize=85mm\epsfbox{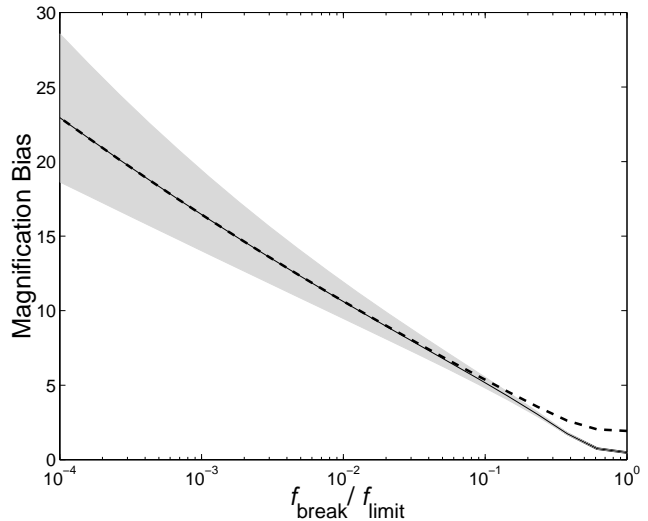}}
\caption {The magnification bias as a function of $f_{break}/f_{limit}$,
for $\alpha_{faint}=1$ (solid line) and $-2$ (dashed line).
The gray area represents the $1\sigma$ error for
the solid-line (i.e., $\alpha$ changes from $-3.09$ to $1$
and from $-3.01$ to $1$).}
\label{MagBias}
\end{figure}

Meanwhile, in order to
obtain a rough estimate of the possible effect of the
magnification bias,
we assumed that below a
flux $f_{break}$, the combined number counts
slope changes from $\alpha=-3.05$ to $\alpha_{faint}$, and we calculated
the magnification bias as function of $f_{break}/f_{limit}$,
for various values of $\alpha_{faint}$,
where $f_{limit}$ is the survey flux limit.
Figure~\ref{MagBias} presents the results for
$\alpha_{faint}=1$ (solid line) and $-2$ (dashed line).
The gray area represents the $1\sigma$ error for
the solid-line (i.e., $\alpha$ changes from $-3.09$ to $1$
and from $-3.01$ to $1$).
Based on these results, for our present rough calculation we adopt
a magnification bias in the range $1-30$.

 From the effects considered, namely, the double magnification bias
and the double flux limit, we find that
for a non-evolving Girardi et al. (1988) mass function,
the predicted fraction of lensed quasars in our survey
is in the range of
$(0.5-4.4)\times 10^{-4} \times 0.66 \times (1-30) \approx (0.3-87)\times10^{-4}$.
Interestingly, this is of the same order as the $95\%$ observational 
upper limit of $3.3\times 10^{-4}$, obtained above.
However, a more thorough calculation needs to be done.
Apart from the unknown properties of FIRST quasars, models
other than SIS (e.g. the NFW model or its generalizations;  Maoz et al. 1997; Wyithe et al. 2000)
may predict a lensing fraction lower or higher by an order of magnitude.
It is possible that, once the detailed properties of the quasars in
our sample are known, such models will predict high rates
for our survey. If so, our results may be able to rule
out some of the cosmological and cluster-structure parameter space.

The FIRST catalog
has many other potential lensing applications
(e.g. Andernach et al. 1998; Helfand et al. 1998b; Lehar et al. 1999).
Additional possibilities are:
(1) to search for lensed galaxies
using the FIRST morphological information;
(2) to combine the FIRST catalog with the good angular resolution of the
2MASS survey in order to find candidates for small-separation
gravitational lensing. We will explore these possibilities in future
papers.

\section*{ACKNOWLEDGMENTS}

We thank M. Brotherton, A. Gal-Yam, and P. GuhaThakurta,
 for observations with the Keck-II
telescope and R. Uglesich for observations with the MDM telescope.
We also thank M. Brotherton for communicating the preliminary results of the FFQS
prior to publication.
We are grateful to the referee, L. Miller, for his useful comments.
EO wishes to thank Orly Gnat for fruitful discussions.
Astronomy at Wise Observatory is supported by grants
from the Israel Science Foundation.
This work is partly supported by grants from NASA and NSF at UCSC and HU (TsK).
This research has made use of the NASA/IPAC Extragalactic Database (NED) which is
operated by the Jet Propulsion Laboratory, California Institute of Technology,
under contract with the National Aeronautics and Space Administration. 
This research has made use of the APM catalogue
run by the Institute of Astronomy in Cambridge.

\end{document}


\pagestyle{empty}
 

\ptlandscape

\begin{deluxetable}{cccrllrrrrcccrll}
\scriptsize
\tablewidth{0pt}
\tablecaption{Lensed Quasar Candidates}
\tablehead{\colhead{\#} & \colhead{$\alpha$} & \colhead{$\delta$} & \colhead{$Sep$} & \colhead{$E$} & \colhead{$O$-$E$} & \colhead{$F^{20cm}$} & \colhead{$E$} & \colhead{$O$-$E$} & \colhead{$F^{20cm}$} & \colhead{$\Delta{m_{I}}$} & \colhead{$\Delta{m_{V}}$} & \colhead{$\Delta{m_{B}}$} & \colhead{$\sigma$} & \multicolumn{2}{c}{Type \& Redshift} \nl
\colhead{} & \colhead{} & \colhead{} & \colhead{} &\multicolumn{3}{c}{West Object} & \multicolumn{3}{c}{East Object} & \colhead{} & \colhead{} & \colhead{} & \colhead{} & \colhead{West} & \colhead{East} \nl
\colhead{} & \multicolumn{2}{c}{J2000} &\colhead{[$''$]} & \colhead{[mag]} &\colhead{[mag]} & \colhead{[mJy]} &\colhead{[mag]} &\colhead{[mag]} & \colhead{[mJy]} &\colhead{[mag]} &\colhead{[mag]} & \colhead{[mag]} & \colhead{} & \colhead{$(T,z)$} & \colhead{$(T,z)$}}
\startdata
 1 & 09:11:06.65 & $+$45:43:45.0 & 22.7 & 15.6       & 2.1 &  6.2 & 17.3 & 2.0 &  1.9 &$-0.91\pm0.06$ &$-0.88\pm0.09$ &$-0.82\pm0.10$ & 0.7    & G,0.098 & G,0.099 \nl
 2 & 09:58:58.57 & $+$29:47:53.9 & 11.4 & 19.2$^{a}$ & 1.3 & 45.8 & 19.2&$-$0.1&138.3 &$+0.35\pm0.16$ &$+0.21\pm0.05$ &$+0.54\pm0.07$ & 3.2    & Q,2.080 & Q,2.75? \nl
 3 & 10:56:10.29 & $+$44:27:16.7 & 24.5 & 19.2       & 0.3 & 49.5 & 19.0 & 0.7 & 55.0 &    \nodata    &  \nodata      &    \nodata    &\nodata & ?,?     & Q,?     \nl
 4 & 11:58:44.66 & $+$41:34:26.3 & 11.7 & 16.7       & 1.8 &  1.1 & 18.1 & 2.5 &  1.0 &$-1.20\pm0.12$ &$-1.07\pm0.11$ &$-1.22\pm0.08$ & 1.1    & G,0.127 & G,0.127 \nl
 5 & 12:03:13.79 & $+$24:21:36.3 & 28.1 & 18.4$^{a}$ & 2.4 & 17.5 & 17.2 & 2.7 &  3.0 &$+0.93\pm0.08$ &$+0.82\pm0.07$ &$+0.79\pm0.14$ & 1.0    & G,0.229 & ?,?     \nl
 6 & 12:06:06.02 & $+$28:55:55.9 &  8.3 & 15.5$^{a}$ & 3.7 &  2.7 & 17.2 & 3.4 &  4.0 &$+0.89\pm0.06$ &$+0.83\pm0.09$ &$^{b}$      & 0.5    & G,0.205 & ?,?     \nl
 7 & 12:44:47.77 & $+$31:39:12.0 & 11.6 & 17.1       & 3.1 &  7.9 & 17.3 & 4.0 &  7.2 &$-0.24\pm0.07$ &$-0.35\pm0.09$ &$-0.68\pm0.35$ & 1.2    & G,0.237 & G,0.241 \nl
 8 & 13:04:25.86 & $+$22:38:22.6 & 28.2 & 16.5$^{a}$ & 3.7 &  3.4 & 17.6 & 2.6 &  1.6 &$-0.85\pm0.04$ &$-0.72\pm0.03$ &$-0.89\pm0.09$ & 1.8    & G,0.223 & G,0.223 \nl
 9 & 13:50:56.42 & $+$50:53:03.6 & 18.5 & 19.0       & 2.7 &  6.6 & 17.2 & 4.4 &  2.6 &$+2.30\pm0.35$ &$+1.87\pm0.27$ &$+0.62\pm0.38$ & 3.2    & ?,?     & ?,?     \nl
10 & 14:43:56.91 & $+$29:19:16.2 & 12.6 & 17.8$^{a}$ & 3.1 &  3.4 & 16.3 & 2.9 & 14.4 &$+0.50\pm0.07$ &$+0.54\pm0.09$ &$+0.51\pm0.24$ & 0.3    & ?,?     & G,0.178 \nl
11 & 14:55:15.47 & $+$30:57:04.4 & 14.8 & 17.2$^{a}$ & 2.1 &  4.2 & 18.2 & 1.7 &  1.4 &$-0.43\pm0.10$ &$-0.33\pm0.07$ &$-0.40\pm0.12$ & 0.7    & G,0.158 & G,0.159 \nl
12 & 15:02:08.29 & $+$25:28:44.9 & 15.9 & 16.7       & 1.4 &  4.1 & 17.4 & 2.7 & 47.1 &$-1.74\pm0.09$ &$-1.90\pm0.09$ &$-2.36\pm0.16$ & 3.2    & G,0.321 & G,0.178 \nl
13 & 16:22:29.54 & $+$35:31:09.1 & 25.7 & 19.8       & 1.7 &  1.7 & 19.9 & 1.7 &218.4 &   \nodata     &  \nodata      &    \nodata    &\nodata & G,?     & Q,?     \nl
14 & 17:08:03.27 & $+$46:48:59.5 & 10.1 & 18.7       & 1.7 &  1.7 & 18.0 & 1.6 &  3.3 &$+0.36\pm0.06$ &$+0.39\pm0.08$ &$+0.37\pm0.09$ & 0.3    & G,0.291 & G,0.291 \nl
15 & 17:20:25.47 & $+$48:03:21.4 & 19.3 & 19.8       & 1.0 & 29.9 & 19.5 & 1.0 &  9.7 &   \nodata     &  \nodata      &    \nodata    &\nodata & ?,?     & Q,?     \nl
\enddata
\tablecomments{Double point source radio/optical objects with
  similar colours ($<3.5 \sigma$).
For each pair,
the coordinates given
  in the table
are for the western object's radio position.
$E$ and $O$ are the APM $E$ and $O$ magnitudes, respectively.
$F^{20cm}$ is the integrated radio flux at $20$ cm.
$\Delta m_{i}$ is the measured $i-$band magnitude difference for
a pair. $\sigma$ is defined in Equation 1.
$(T,z)$ is the type (Q-Quasar; G-Galaxy; ?-Unknown)
and redshift of the objects.
\newline
$^{a}$ - The APM magnitude for at least one band is not known,
and therefore the $E$ and $O$ magnitudes are taken from the USNO-A1.0 catalog.
$^{b}$ - Close to bright star, difficult to observe in the $B$ band.
\\
{\bf Notes on individual objects:}
\\
$2$ - A red spectrum (not shown) of the eastern object shows $CIII]\lambda 1909$
emission at $7125$\AA.
\\
$3$ - Unidentified narrow line at $4305$\AA ~present only in the western
object, and unidentified broad emission feature
at $4914$\AA ~present only in the eastern object.
\\
$5$ - No lines are present in the spectrum of the eastern object.
\\
$6$ - No emission lines, Ca $H\&K$ is probably present in the western source.
\\
$8$ - No emission lines, redshift based on Ca $H\&K$ break.
\\
$9$ - No lines in the spectra of either members, but there is a clear difference in continuum shapes redward of $4500$\AA.
\\
$10$ - $[OII]$, $H\alpha$, $[NII]$ and Ca $H\&K$ in the eastern object, no lines in the western source.
\\
$12$ - Strong emission lines: $[OII]$, $H\alpha$, $[OIII]$ and $H\beta$ in the western object, no lines in the eastern source.
\\
$13$ - Eastern object has a narrow emission line at $4904$\AA, and the
western object a broad emission line at $5389$\AA.
\\
$14$ - Interacting galaxies based on Keck-II images.
\\
$15$ - Broad emission feature in the eastern object at $4296$\AA.}
\end{deluxetable}